\documentclass[preprint,aps,  nofootinbib]{revtex4}
\usepackage{amssymb}
\usepackage{amsmath}
\usepackage{graphicx}

\newcommand\beq{ \begin{eqnarray} }
\newcommand\eeq{ \end{eqnarray} }

\begin{document}

\title{Shear Viscosity to Entropy Density Ratio of QCD below the
Deconfinement Temperature}
\author{Jiunn-Wei Chen}
\author{Eiji Nakano}
\affiliation{Department of Physics, National Taiwan University, Taipei 10617, Taiwan}

\begin{abstract}
Using chiral perturbation theory we investigate the QCD\ shear viscosity ($%
\eta $) to entropy density ($s$) ratio below the deconfinement temperature ($%
\sim 170$ MeV) with zero baryon number density. It is found that $\eta /s$
of QCD is monotonically decreasing in temperature ($T)$ and reaches $0.6$
with estimated $\sim 50\%$ uncertainty at $T=120$ MeV. A naive extrapolation
of the leading order result shows that $\eta /s$ reaches the $1/4\pi $
minimum bound proposed by Kovtun, Son, and Starinets using string theory
methods at $T\sim 200$ MeV. This suggests a phase transition or cross over
might occur at $T\lesssim 200$ MeV in order for the bound to remain valid.
Also, it is natural for $\eta /s$ to stay close to the minimum bound around
the phase transition temperature as was recently found in heavy ion
collisions.
\end{abstract}

\maketitle

%\date{\today}

\section{Introduction}

Shear viscosity $\eta $ characterizes how strongly particles interact and
move collectively in a many body system. In general, strongly interacting
systems have smaller $\eta $ than the weakly interacting ones. This is
because $\eta $ is proportional to $\epsilon \tau _{mft}$, where $\epsilon $
is the energy density and $\tau _{mft}$ is the mean free time, which is
inversely proportional to particle scattering cross section. Recently a
universal minimum bound for the ratio of $\eta $ to entropy density $s$ was
proposed by Kovtun, Son, and Starinets \cite{KOVT1}. The bound, 
\begin{equation}
\frac{\eta }{s}\geq \frac{1}{4\pi }\ ,
\end{equation}%
is found to be saturated for a large class of strongly interacting quantum
field theories whose dual descriptions in string theory involve black holes
in anti-de Sitter space \cite%
{Policastro:2001yc,Policastro:2002se,Herzog:2002fn,Buchel:2003tz}.

Recently, $\eta /s$ close to the minimum bound were found in relativistic
heavy ion collisions (RHIC) \cite{RHIC,Molnar:2001ux,Teaney:2003pb}. This
discovery came as a surprise. Traditionally, quark gluon plasma (QGP)---the
phase of QCD above the deconfinement temperature $T_{c}$($\sim 170$ MeV at
zero baryon density \cite{KL04})---was thought to be weakly interacting.
Partly because lattice QCD simulations of the QGP equation of state above $%
2T_{c}$ were not inconsistent with that of an ideal gas of massless
particles, $e=3p$, where $e$ is the the energy density and $p$ is the
pressure of the system \cite{KL04}. However, recent analyses of the elliptic
flow generated by non-central collisions in RHIC \cite%
{Molnar:2001ux,Teaney:2003pb} and lattice simulations of a gluon plasma \cite%
{Nakamura:2004sy} yielded $\eta /s$ close to the the minimum bound at just
above $T_{c}$. This suggests QGP is strongly interacting at this temperature.%
\footnote{%
See also \cite{Hatsuda03,Datta03,Umeda02}. For discussions of the possible
microscopic stucture of such a state, see \cite%
{Shuryak:2004tx,Koch:2005vg,Liao:2005pa,GerryEd,GerryRho}} (However, see
Ref. \cite{Asakawa:2006tc} for a\ different interpretation.)

Given this situation, one naturally wonders if $\eta /s$ of QCD was already
close to the minimum bound at just above $T_{c}$, what would happen if we
keep reducing the temperature such that the coupling constant of QCD gets
even stronger? Will the $\eta /s$ minimum bound hold up below $T_{c}$? If
the bound does hold up, what is the mechanism? Is the change of degrees of
freedom through a phase transition or cross over sufficient to save the
bound? If\ the bound does not hold up, what is the implication to string
theory?

To explore these issues, we use chiral perturbation theory ($\chi $PT) and
the linearized Boltzmann equation to perform a model independent calculation
to the $\eta /s$ of QCD in the confinement phase. Earlier attempts to
compute meson matter viscosity using the Boltzmann equation and
phenomenological phase shifts in the context of RHIC hydrodynamical
evolution after freeze out can be found in Refs. \cite{Davesne,DOBA1,DOBA2}.
In the deconfinement phase, state of the art perturbative QCD calculations
of $\eta $ can be found in Refs. \cite{Arnold:2003zc,Arnold:2000dr}.

\section{Linearized Boltzmann Equation for Low Energy QCD}

In the hadronic phase of QCD with zero baryon-number density, the dominant
degrees of freedom are the lightest hadrons---the pions. The pion mass $%
m_{\pi }=139$ MeV is much lighter than the mass of the next lightest
hadron---the kaon whose mass is $495$ MeV. Given that $T_{c}$ is only $\sim
170$ MeV, it is sufficient to just consider the pions in the calculation of
thermodynamical quantities and transport coefficients for $T\ll T_{c}$.

The interaction between pions can be described by chiral perturbation theory
($\chi $PT) in a systematic expansion in energy and quark ($u$ and $d$
quark) masses \cite{ChPT,GL,Colangelo:2001df}. $\chi $PT is a low energy
effective field theory of QCD. It describes pions as Nambu-Goldstone bosons
of the spontaneously broken chiral symmetry. At $T\ll T_{c}$, the
temperature dependence in $\pi \pi $ scattering can be calculated
systematically. At $T=T_{c}$, however, the theory breaks down due to the
restoration of chiral symmetry.\footnote{%
The QCD chiral restoration temperature and the deconfinement temperature
happen to be close to each other at zero baryon density. We do not
distinguish the two in this paper.}

The shear viscosity $\eta $ of the pion gas can be calculated either using
the Boltzmann equation or the Kubo formula. Since the Boltzmann equation
requires semi-classical descriptions of particles with definite position,
energy and momentum except during brief collisions, the mean free path is
required to be much greater than the range of interaction. Thus the
Boltzmann equation is usually limited to low temperature systems. The Kubo
formula does not have this restriction. In this approach $\eta $ can be
calculated through the linearized response function

\begin{equation}
\eta =-\frac{1}{5}\int_{-\infty }^{0}\mathrm{d}t^{\prime }\int_{-\infty
}^{t^{\prime }}\mathrm{d}t\int \mathrm{d}x^{3}\langle \left[
T^{ij}(0),T^{ij}(\mathbf{x},t)\right] \rangle
\end{equation}%
with $T^{ij}$ the spacial part of the off-diagonal energy momentum tensor.
One might think a perturbative calculation of the above two point function
will give the answer for $\eta $. But this can not be true if $\eta
\varpropto \tau _{mft}$, as mentioned above, for $\tau _{mft}\rightarrow
\infty $ in the free case. Indeed, the Kubo formula involves an infinite
number of diagrams at the leading order (LO) \cite{Jeon}. However, in a weak
coupling $\phi ^{4}$ theory, it is proven that the summation of LO diagrams
is equivalent to solving the linearized Boltzmann equation with temperature
dependent particle masses and scattering amplitudes \cite{Jeon}. This proof
extended the applicable range of the Boltzmann equation to higher
temperature but is restricted to weak coupling theories. In the case we are
interested (QCD with $T<140$ MeV), the pion mean free path is always greater
than the range of interaction ($\sim 1$ fm) by a factor of $10^{3}$. Thus,
even though the coupling in $\chi $PT is too strong to use the result of
Ref. \cite{Jeon}, the temperature is still low enough that the use of the
Boltzmann equation is justified.

The Boltzmann equation describes the evolution of the isospin averaged pion
distribution function $f=f(\mathbf{x},\mathbf{p},t)\equiv f_{p}(x)$ (a
function of space, time and momentum) as 
\begin{equation}
\frac{p^{\mu }}{E_{p}}\partial _{\mu }f_{p}(x)=\frac{g_{\pi }}{2}%
\int_{123}d\Gamma _{12;3p}\left\{
f_{1}f_{2}(1+f_{3})(1+f_{p})-(1+f_{1})(1+f_{2})f_{3}f_{p}\right\} \ ,
\end{equation}%
where $E_{p}=\sqrt{\mathbf{p}^{2}+m_{\pi }^{2}}$and $g_{\pi }=3$ is the
degeneracy factor for three pions , 
\begin{equation}
d\Gamma _{12;3p}\equiv \frac{1}{2E_{p}}|\mathcal{T}|^{2}\prod_{i=1}^{3}\frac{%
d^{3}\mathbf{k}_{i}}{(2\pi )^{3}(2E_{i})}\times (2\pi )^{4}\delta
^{4}(k_{1}+k_{2}-k_{3}-p)\ ,
\end{equation}%
and where $\mathcal{T}$ is the scattering amplitude for particles with
momenta $1,2\rightarrow 3,p$. In $\chi $PT, the LO isospin averaged $\pi \pi 
$ scattering amplitude in terms of Mandelstam variables ($s,t$, and $u$) is 
\begin{equation}
|\mathcal{T}|^{2}=\frac{1}{9}\sum_{I=0,1,2}(2I+1)|\mathcal{T}^{(I)}|^{2}=%
\frac{1}{9f_{\pi }^{4}}\left\{ 21m_{\pi }^{4}+9s^{2}-24M_{\pi
}^{2}s+3(t-u)^{2}\right\} \ .
\end{equation}%
The temperature dependence in pion mass and pion scattering amplitudes can
be treated as higher order corrections.

In local thermal equilibrium, the distribution function $f_{p}^{(0)}(x)=%
\left( e^{\beta (x)V_{\mu }(x)p^{\mu }}-1\right) ^{-1}$ with $\beta (x)$ the
inverse temperature and $V^{\mu }(x)$ the four velocity at the space-time
point $x$. A small deviation of $f_{p}$ from local equilibrium is
parametrized as 
\begin{equation}
f_{p}(x)=f_{p}^{(0)}(x)\left[ 1-\left\{ 1+f_{p}^{(0)}(x)\right\} \chi _{p}(x)%
\right] \ ,  \label{df0}
\end{equation}%
while the energy momentum tensor is%
\begin{equation}
T_{\mu \nu }(x)=g_{\pi }\int \frac{\mathrm{d}^{3}\mathbf{p}}{(2\pi )^{3}}%
\frac{p_{\mu }p_{\nu }}{E_{p}}f_{p}(x)\ .  \label{dT}
\end{equation}%
We will choose the $\mathbf{V}(x)=0$ frame for the point $x$. This implies $%
\partial _{\nu }V^{0}=0$ after taking a derivative on $V_{\mu }(x)V^{\mu
}(x)=1$. Furthermore, the conservation law at equilibrium $\partial _{\mu
}T^{\mu \nu }|_{\chi _{p}=0}=0$ allows us to replace $\partial _{t}\beta (x)$
and $\partial _{t}\mathbf{V}(x)$ by terms proportional to $\nabla \cdot 
\mathbf{V}(x)$ and $\mathbf{\nabla }\beta (x)$. Thus, to the first order in
a derivative expansion, $\chi _{p}(x)$ can be parametrized as 
\begin{equation}
\chi _{p}(x)=\beta (x)A(p)\nabla \cdot \mathbf{V}(x)+\beta (x)B(p)\left( 
\hat{p}_{i}\hat{p}_{j}-\frac{1}{3}\delta _{ij}\right) \left( \frac{\nabla
_{i}V_{j}(x)+\nabla _{j}V_{i}(x)}{2}-\frac{1}{3}\delta _{ij}\nabla \cdot 
\mathbf{V}(x)\right) \ ,  \label{df1}
\end{equation}%
where $i$ and $j$ are spacial indexes.\footnote{%
A non-derivative term is not allowed since $f_{p}$ should be reduced to $%
f_{p}^{(0)}$ when $\beta $ and $V^{\mu }$ become independent of $x$. There
is no term with single spacial derivative on $\beta (x)$ either. The only
possible term $\left( \mathbf{V}\cdot \mathbf{\nabla }\right) \beta (x)$
vanishes in the $\mathbf{V}(x)=0$ frame.} $A$ and $B$ are functions of $x$
and $p$. But we have suppressed the $x$ dependence.

Substituting (\ref{df1}) into the Boltzmann equation, one obtains a
linearized equation for $B$ 
\begin{eqnarray}
\left( p_{i}p_{j}-\frac{1}{3}\delta _{ij}\mathbf{p}^{2}\right) &=&\frac{%
g_{\pi }E_{p}}{2}\int_{123}d\Gamma
_{12;3p}(1+n_{1})(1+n_{2})n_{3}(1+n_{p})^{-1}  \notag \\
&&\times \left[ B_{ij}(p)+B_{ij}(k_{3})-B_{ij}(k_{2})-B_{ij}(k_{1})\right]
\equiv g_{\pi }\hat{F}_{ij}\left[ B\right] \ ,  \label{LB1} \\
B_{ij}(p) &\equiv &B(p)\left( \hat{p}_{i}\hat{p}_{j}-\frac{1}{3}\delta
_{ij}\right) \ ,  \notag
\end{eqnarray}%
where we have dropped the factor $\left( \nabla _{i}V_{j}(x)+\nabla
_{j}V_{i}(x)-\text{trace}\right) $ contracting both sides of the equation
and write $f_{i}^{(0)}(x)$ at this point as $n_{i}=\left( e^{\beta
E_{i}}-1\right) ^{-1}$. There is another integral equation associated with $%
\nabla \cdot \mathbf{V}(x)$ which is related to the bulk viscosity $\zeta $
that will not be discussed in this paper. The $\mathbf{\nabla }\cdot \beta $
and $\partial _{t}\mathbf{V}$ terms in $p^{\mu }\partial _{\mu }f_{p}^{(0)}$
will cancel each other by the energy momentum conservation in equilibrium
mentioned above.

In equilibrium the energy momentum tensor depends on pressure ${\mathcal{P}}%
(x)$ and energy density $\epsilon (x)$ as $T_{\mu \nu }^{(0)}(x)=\left\{ {%
\mathcal{P}}(x)+\epsilon (x)\right\} V_{\mu }(x)V_{\nu }(x)-{\mathcal{P}}%
(x)\delta _{\mu \nu }$. A small deviation away from equilibrium gives
additional contribution to $T_{\mu \nu }$ whose spacial components define
the shear and bulk viscosity 
\begin{equation}
\delta T_{ij}=-2\eta \left( \frac{\nabla _{i}V_{j}(x)+\nabla _{j}V_{i}(x)}{2}%
-\frac{1}{3}\delta _{ij}\nabla \cdot \mathbf{V}(x)\right) +\zeta \delta
_{ij}\nabla \cdot \mathbf{V}(x)\ .
\end{equation}%
After putting everything together we obtain 
\begin{eqnarray}
\eta &=&\frac{g_{\pi }\beta }{10}\int \frac{\mathrm{d}^{3}\mathbf{p}}{(2\pi
)^{3}}\frac{1}{E_{p}}n_{p}\left( 1+n_{p}\right) B_{ij}(p)\left( p_{i}p_{j}-%
\frac{1}{3}\delta _{ij}\mathbf{p}^{2}\right) \   \notag \\
&=&\frac{g_{\pi }^{2}\beta }{10}\int \frac{\mathrm{d}^{3}\mathbf{p}}{(2\pi
)^{3}}\frac{1}{E_{p}}n_{p}\left( 1+n_{p}\right) B_{ij}(p)\hat{F}_{ij}\left[ B%
\right] \equiv g_{\pi }^{2}\langle B|\hat{F}[B]\rangle \ .  \label{EQ1}
\end{eqnarray}%
Here we see immediately that if the scattering cross section is scaled by a
factor $\alpha $, 
\begin{equation}
d\Gamma _{12;3p}\rightarrow \alpha \left( d\Gamma _{12;3p}\right) \ ,
\label{s1}
\end{equation}%
then Eqs. (\ref{LB1}) and (\ref{EQ1}) imply the following scaling 
\begin{eqnarray}
B_{ij}(p) &\rightarrow &\alpha ^{-1}B_{ij}(p)\ ,  \notag \\
\eta &\rightarrow &\alpha ^{-1}\eta \ ,  \label{s2}
\end{eqnarray}%
with $\eta $ proportional to the inverse of scattering cross-section. This
non-perturbative result is a general feature for the linearized Boltzmann
equation with two-body elastic scattering.

To find a solution for $B(p)$, one can just solve Eq. (\ref{LB1}). But here
we follow the approach outlined in Ref. \cite{DOBA1,DOBA2} to assume that $%
B(p)$ is a\ smooth function which can be expanded using a specific set of
orthogonal polynomials 
\begin{equation}
B(p)=g_{\pi }^{-1}|\mathbf{p}|^{y}\sum_{r=0}^{\infty }b_{r}B^{(r)}(z(p)),
\label{BP1}
\end{equation}%
where $B^{(r)}(z)$ is a polynomial up to $z^{r}$ and $b_{r}$ is its
coefficient. The overall factor $|\mathbf{p}|^{y}$ will be chosen by trial
and error to get the fastest convergence. The orthogonality condition

\begin{equation}
\int \frac{\mathrm{d}^{3}\mathbf{p}}{(2\pi )^{3}}\frac{\mathbf{p}^{2}}{E_{p}}%
n_{p}\left( 1+n_{p}\right) |\mathbf{p}|^{y}B^{(r)}(z)B^{(s)}(z)\propto
\delta _{r,s}\   \label{BP2}
\end{equation}%
can be used to construct the $B^{(r)}(z)$ polynomials up to normalization
constants. For simplicity, we will choose 
\begin{equation*}
B^{(0)}(z)=1\ .
\end{equation*}

With this expansion, the consistency condition for $B(p)$ in Eq.(\ref{EQ1})
yields%
\begin{equation}
\eta =\sum_{r}b_{r}L^{(r)}=\sum_{r,s}b_{r}\langle B^{(r)}|\hat{F}%
[B^{(s)}]\rangle b_{s}\ ,  \label{ME1}
\end{equation}%
where 
\begin{equation}
L^{(r)}=\frac{\beta }{15}\int \frac{\mathrm{d}^{3}\mathbf{p}}{(2\pi )^{3}}%
\frac{\mathbf{p}^{2}}{E_{p}}n_{p}\left( 1+n_{p}\right) |\mathbf{p}%
|^{y}B^{(r)}(p)\propto \delta _{0,r}\ .
\end{equation}%
Since $b_{r}$ is a function of $m_{\pi }$, $f_{\pi }$ and $T$, the $b_{r}$'s
in Eq.(\ref{ME1}) are in general independent functions, such that $%
L^{(r)}=\sum_{s}\langle B^{(r)}|\hat{F}[B^{(s)}]\rangle b_{s}$ [one can show
that this solution of $b_{s}$ gives a unique solution of $\eta $], or
equivalently

\begin{equation}
\delta _{0,r}L^{(0)}=\sum_{s}\langle B^{(r)}|\hat{F}[B^{(s)}]\rangle b_{s}\ .
\label{Lr}
\end{equation}%
This will allow us to solve for the $b_{s}$ and obtain the shear viscosity 
\begin{equation}
\eta =b_{0}L^{(0)}.  \label{ETA2}
\end{equation}

In the next section, we will show that this expansion converges well, such
that one does not need to keep many terms on the right hand side of Eq.(\ref%
{Lr}). If only the $s=0$ term was kept, then 
\begin{equation}
\eta \simeq \frac{\left( L^{(0)}\right) ^{2}}{\langle B^{(0)}|\hat{F}%
[B^{(0)}]\rangle }\ .
\end{equation}

The calculation of the entropy density $s$ is more straightforward since $s$%
, unlike $\eta $, does not diverge in a free theory. In $\chi $PT, the
interaction contributions are all higher order in our LO calculation. Thus
we just compute the $s$ for a free pion gas: 
\begin{equation}
s=-g_{\pi }\beta ^{2}\frac{\partial }{\partial \beta }\frac{\log {Z}}{\beta }%
\ ,  \label{S}
\end{equation}%
where the partition function ${Z}$ for free pions is 
\begin{equation}
\frac{\log {Z}}{\beta }=-\frac{1}{\beta }\int \frac{\mathrm{d}^{3}\mathbf{p}%
}{(2\pi )^{3}}\log \left\{ 1-e^{-\beta E(p)}\right\} \ ,
\end{equation}%
up to temperature independent terms. \ 

\section{Numerical Results}

%------------------------------------------------------
\begin{figure}[tbp]
\begin{center}
\includegraphics[height=8cm]{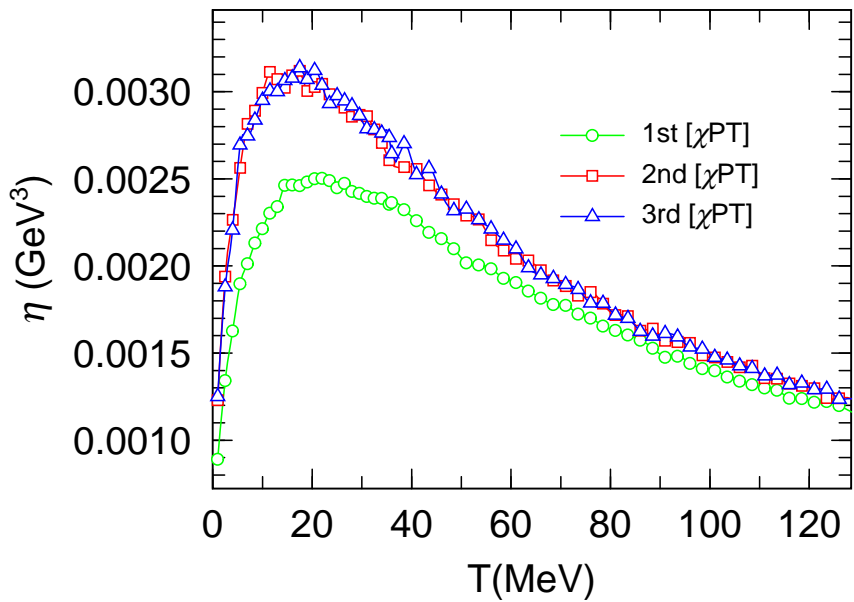}
\end{center}
\caption{(Color online) Shear viscosity as a function of temperature in LO $%
\protect\chi $PT. $1_{\mathrm{st}},$ $2_{\mathrm{nd}}$ and $3_{\mathrm{rd}}$
are the results of keeping the first one, two and three polynomials on the
right hand side of Eq.(\protect\ref{Lr}), respectively.}
\end{figure}
%------------------------------------------------------  

In this section we present the results for $\eta $ and $\eta /s$ of QCD up
to $T=120$ MeV at zero baryon number density. In Fig. 1 the LO $\chi $PT
result of $\eta $ using the linearized Boltzmann equation is shown. The
lines with circles, squares and triangles correspond to keeping the first
one, two and three polynomials on the right hand side of Eq.(\ref{Lr}),
respectively. We have used $y=0$ and $z(p)=|\mathbf{p}|$ to construct the
polynomials. The figure shows the expansion converges rapidly. As a test of
the calculation, we also reproduce the shear viscosity result of Ref. \cite%
{Jeon} for $\phi ^{4}$ theory by setting the scattering amplitude $\mathcal{T%
}=\lambda $ to be a constant. In $\phi ^{4}$ theory, $\eta $ is
monotonically increasing in $T$. If $T\gg m_{\phi }$, $\eta $ $\propto
T^{3}/\lambda ^{2}$ with $T^{3}$ given by dimensional analysis and $\lambda
^{-2}$ by the scaling of coupling shown in Eqs. (\ref{s1}) and (\ref{s2}).
In $\chi $PT, however, $\eta $ is not monotonic in $T$. At $T\ll m_{\pi }$,
the scattering amplitude is close to a constant, thus $\chi $PT behaves like
a $\phi ^{4}$ theory. But at $T\gg m_{\pi }$, $\mathcal{T}$ $\propto
T^{2}/f_{\pi }^{2}$ and $\eta $ $\propto f_{\pi }^{4}/T$. At what
temperature the transition from $\eta $ $\propto T^{3}$ to $\eta $ $\propto
1/T$ takes place depends on the detail of dynamics. In $\chi $PT, this
temperature is around $20$ MeV.

The radius of convergence in momentum for $\chi $PT is typically $4\pi
f_{\pi }\sim 1$ GeV. To translate this radius of convergence into
temperature, we compute the averaged center of mass momentum $\langle |%
\mathbf{p}|\rangle =\sqrt{\langle B|\mathbf{p}^{2}|\hat{F}[B]\rangle
/\langle B|\hat{F}[B]\rangle }$. We found that for $T=120$ and $140$ MeV, $%
\langle |\mathbf{p}|\rangle \simeq 460$ and $530$ MeV $<4\pi f_{\pi }$.
However, $\chi $PT is expected to break down at the chiral restoration
temperature ($\sim 170$ MeV). Thus our LO\ $\chi $PT result can only be
trusted up to $T\sim 120$ MeV. At the next-to-leading order (NLO), it is
known that the isoscalar $\pi \pi $ scattering length will be increased by $%
\sim 40\%$ \cite{Colangelo:2001df}. This will increase the cross section by $%
\sim 100\%$ and reduce $\eta $ by $\sim 50\%$ near threshold. This is an
unusually large NLO correction since a typical NLO correction at threshold
is $\lesssim 20\%$. The large chiral corrections does not persist at the
higher order. At the next-to-next-to-leading order (NNLO), the correction is 
$\sim 10\%$ \cite{Colangelo:2001df}. Thus, to compute $\eta $ to $10\%$
accuracy, a NLO $\chi $PT calculation is needed.

\begin{figure}[tbp]
\begin{center}
\includegraphics[height=8cm]{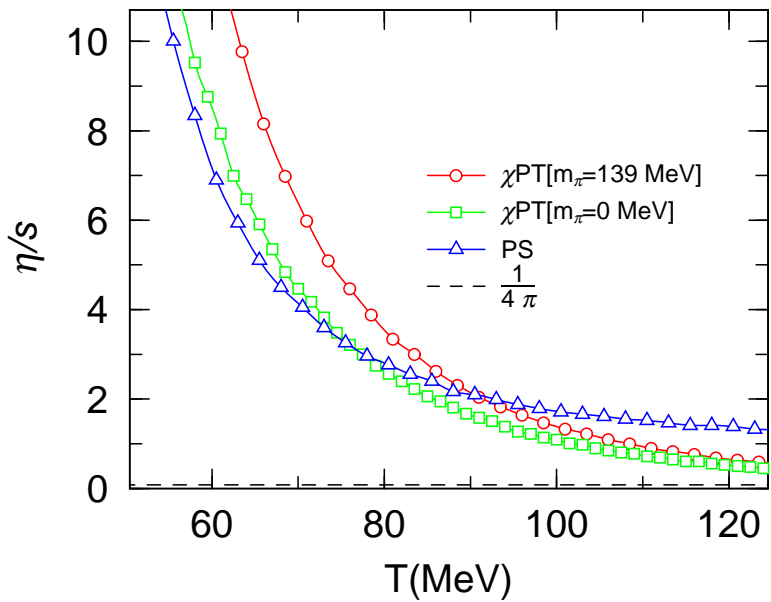}
\end{center}
\caption{(Color online) Shear viscosity to entropy density ratios as
functions of temperature. Line with circles (rectangles) is the LO $\protect%
\chi $PT result with $m_{\protect\pi }=139\,(0)$ MeV and $f_{\protect\pi %
}=93(87)$ MeV. Line with triangles is the result using $\protect\pi \protect%
\pi $ phase shifts (PS). Dashed line is the conjectured minimum bound $1/4%
\protect\pi \simeq 0.08$.}
\end{figure}

The LO $\chi $PT result for $\eta /s$ is shown in Fig. 2 (line with
rectangles). The error is estimated to be $\sim 50\%$ up to 120 MeV. $\eta
/s $ is monotonically decreasing and reaches $0.6$ at $T=120$ MeV.\ This is
similar to the behavior in the $m_{\pi }=0$ case (shown as the line with
rectangles) where $\eta /s\propto f_{\pi }^{4}/T^{4}$ with $s\propto T^{3}$
from dimensional analysis and $f_{\pi }=87$ MeV in the chiral limit \cite%
{GL,Colangelo:2001df}.

For comparison, we also show the result using phenomenological $\pi \pi $
phase shifts \cite{Schenk:1991xe} for $\eta $ but free pions for $s$. (Our
result for $\eta $ is in good agreement with that of \cite{DOBA2} for $T$
between $60$ and $120$ MeV. For an earlier calculation using the
Chapman-Enskog approximation, see Ref. \cite{Prakash:1993kd}.) This amounts
to take into account part of the NLO $\pi \pi $ scattering effects but
ignore its temperature dependence and the interaction in $s$. Since not all
the NLO effects are accounted for, this $\eta /s$ is not necessarily more
accurate than the one using LO $\chi $PT. The comparison, however, gives us
some feeling of the size of error for the LO result we present here. Thus,
an error of $\sim 100\%$ at $T=120$ MeV for the LO result might be more
realistic.

Naive extrapolations of the three $\eta /s$ curves show that the $1/4\pi
=0.08$ minimum bound conjectured from string theory might never be reached
as in phase shift result (the first scenario), or more interestingly, be
reached at $T\sim 200$ MeV, as in the LO $\chi $PT result (the second
scenario). In both scenarios, we see no sign of violation of the universal
minimum bound for $\eta /s$ below $T_{c}$. \ But to really make sure the
bound is valid from $120$ MeV to $T_{c}$, a lattice computation as was
performed to gluon plasma above $T_{c}$ \cite{Nakamura:2004sy} is needed. In
the second scenario, assuming the bound is valid for QCD, then either a
phase transition or cross over should occur before the minimum bound is
reached at $T\sim 200$ MeV. Also, in this scenario, it seems natural for $%
\eta /s$ to stay close to the minimum bound around $T_{c}$ as was recently
found in heavy ion collisions.

In the second scenario, one might argue that the existence of phase
transition is already known, otherwise we will not have spontaneous symmetry
breaking and the corresponding Nambu-Golstone boson theory at low
temperature in the first place. Indeed, it is true in the case of QCD.
However, if the $\eta /s$ bound is really set by Nature, then a phase
transition is inevitable in the vicinity of the temperature where the bound
is reached. For a spontaneous symmetry breaking theory, the general feature
of $\eta /s$ we see here seems generic. At very high $T$, collective motion
is weak, thus $\eta /s$ gets smaller at lower $T$. At very low $T$ in the
symmetry breaking phase, the Nambu-Goldstone bosons are weakly interaction
at low temperature, thus $\eta /s$ gets smaller at higher $T$. \ A phase
transition should occur before the extrapolated $\eta /s$ curve coming from
high $T$ reaches the bound at $T_{1}$. Similarly, a phase transition should
occur before the extrapolated $\eta /s$ curve coming from low $T$ reaches
the bound at $T_{2}$. Thus the range of phase transition is $T_{1}\leq
T_{c}\leq T_{2}$. However, it is also possible that the first scenario takes
place and $\eta /s$ bounces back to higher values without a phase
transition. In this case, it is less clear what makes $\eta /s$
non-monotonic and it certainly deserves further study.

It is interesting to note that the degeneracy factor $g_{\pi }$ drops out of 
$\eta $ while the entropy $s$ is proportional to $g_{\pi }$ as in Eqs.(\ref%
{ME1}) and (\ref{S}), respectively. This suggests the $\eta /s$ bound might
be violated if a system has a large particle degeneracy factor.\footnote{%
We thank Thomas Cohen for pointing this out to us. This possibility was also
mentioned in Ref. \cite{KOVT1}.} For QCD, large $g_{\pi }$ can be obtained
by having a large number of quark flavors $N_{f}$ with $g_{\pi }\sim
N_{f}^{2}$. However, the existence of confinement demands that the number of
colors $N_{c}$ should be of order $N_{f}$ to have a negative QCD beta
function. After using $f_{\pi }\propto \sqrt{N_{c}}$, the combined $N_{c}$
and $N_{f}$ scaling of $\eta /s$ is 
\begin{equation}
\frac{\eta }{s}\propto \frac{f_{\pi }^{4}}{g_{\pi }T^{4}}\propto \frac{%
N_{c}^{2}}{N_{f}^{2}}\ ,
\end{equation}%
which is of order one. Thus QCD with large $N_{c}$ and $N_{f}$ can still be
consistent with the $\eta /s$ bound below $T_{c}$.

%%%%%%%%%%%%%%%%%%%%%%%%%%%%%%%%%%%%%%%%%%%%%%%%%%%%%%%%%%

\section{Conclusion}

We have explored whether the conjectured $\eta /s$ minimum bound will hold
up below the QCD deconfinement temperature. Using chiral perturbation theory
and the linearized Boltzmann equation we have computed the QCD $\eta /s$
ratio at zero baryon number density and for $T\leq 120$ MeV. It is found
that $\eta /s$ is monotonic decreasing in $T$ and it reaches $0.6$ with
estimated $50\%$ uncertainty at $T=120$ MeV. Naive extrapolations have shown
that $\eta /s$ met the $1/4\pi $ minimum bound conjectured from string
theory at $T\sim 200$ MeV as in the LO $\chi $PT case, or $\eta /s$ stayed
above the bound as in the phenomenological phase shift case. In the former
case, in order for the $\eta /s$ lower bound to remain valid at higher
temperature, a phase transition or cross over should occur at $T\lesssim 200$
MeV before the bound is reached. We argued that this might be a general
feature for spontaneous symmetry breaking theories that the extrapolation of
the low(high) temperature $\eta /s$ curve sets a upper(lower) bound on $%
T_{c} $. Our result also suggests that it is natural for $\eta /s$ to stay
close to the lower bound around the phase transition temperature as was
recently found in heavy ion collisions.

As this paper was being finished, reference \cite{Csernai:2006zz} appeared.
In that paper, some of the relations between $T_{c}$ and the $\eta /s$ bound
are also discussed.

\section{Acknowledgements}

We would like to thank Thomas Cohen, Will Detmold, Chung-Wen Kao and D.T.
Son for useful discussions. EN is grateful to Kyosuke Tsumura for valuable
comments on viscosity in QGP. This work was supported by the NSC and NCTS of
Taiwan, ROC.

%%%%%%%%%%%%%%%%%%%%%%%%%%%%%%%%%%%%%%%%%%%%%%%%%%%%%%%%%%%%%%

\end{document}